\documentstyle[aps,prl,twocolumn,psfig]{revtex}
\begin{document}
\preprint{\vbox{\hbox{DPNU-00-xx}}}
\def\vev#1{\langle #1 \rangle}
\def\dfrac{\displaystyle\frac}
\def\eqlab#1{\label{eqn:#1}}

\def\eqref#1{Eq.(\ref{eqn:#1})}
\def\Eqref#1{Equation~(\ref{eqn:#1})}
\def\eqsref#1{Eqs.(\ref{eqn:#1})}
\def\Eqsref#1{Equations~(\ref{eqn:#1})}
\def\eqvref#1{(\ref{eqn:#1})}
\def\Eqvref#1{(\ref{eqn:#1})}
\def\mz{m_Z^{}}
\def\mh{m_h^{}}
\def\PR#1#2#3{Phys. Rev. {\bf #1} (#3) #2 }
\def\PRL#1#2#3{Phys. Rev. Lett. {\bf #1} (#3) #2 }
\def\PL#1#2#3{Phys. Lett. {\bf #1} (#3) #2 }
\def\NL#1#2#3{Nucl. Phys. {\bf #1} (#3) #2 }
\def\NP#1#2#3{Nucl. Phys. {\bf #1} (#3) #2 }
\def\PREP#1#2#3{Phys. Report {\bf #1} (#3) #2 }
\def\Mod#1#2#3{Mod. Phys. Lett. {\bf #1} (#3) #2 }
\def\PTP#1#2#3{Prog. Theor. Phys. {\bf #1} (#3)#2 }
\def\EPJ#1#2#3{Eur. Phys. J. {\bf #1} (#3) #2 }
\def\pl#1#2#3{{\it Phys. Lett. }{\bf B#1~}(#2)~#3}
\def\zp#1#2#3{{\it Z. Phys. }{\bf C#1~}(#2)~#3}
\def\prl#1#2#3{{\it Phys. Rev. Lett. }{\bf #1~}(#2)~#3}
\def\rmp#1#2#3{{\it Rev. Mod. Phys. }{\bf #1~}(#2)~#3}
\def\prep#1#2#3{{\it Phys. Rep. }{\bf #1~}(#2)~#3}
\def\pr#1#2#3{{\it Phys. Rev. }{\bf D#1~}(#2)~#3}
\def\np#1#2#3{{\it Nucl. Phys. }{\bf B#1~}(#2)~#3}
\def\d#1{\left[ #1 \right]_D}
\def\f#1{\left[ #1 \right]_F}
\def\a#1{\left[ #1 \right]_A}

\setcounter{footnote}{0}
\setcounter{page}{1}
\setcounter{section}{0}
\setcounter{subsection}{0}
\setcounter{subsubsection}{0}

\noindent
{\large{\bf Comment on ``Linking solar and long baseline
terrestrial neutrino experiments''}}\\

\par
In the recent letter \cite{ABR},
 it has been said that
 the measurements of $U_{e3}$
 would help discriminate between
 the possible solar neutrino solutions under the
 natural conditions with the neutrino mass hierarchies of
 $m_1 \ll m_2 \ll m_3$ and $m_1 \sim m_2 \gg m_3$, where
 $m_i$ is the $i$-th generation neutrino absolute mass.
However, in general the value of $U_{e3}$ cannot predict
 the solar neutrino solutions
 without one additional nontrivial
 condition as the following reasons\cite{HS}.
Neglecting the $CP$ phases in the lepton sector,
 the number of independent parameters in
 the Majorana mass matrix of neutrino
 are six.
Five parameters are enough to determine the MNS
 matrix, since overall factor in the
 neutrino mass matrix does not contribute to
 the MNS matrix.
Thus we need five input parameters
 in order to determine the MNS matrix,
 and its element $U_{e3}$.
Since the neutrino oscillation experiments except for
 the CHOOZ
 give us only four input parameters
 $\Delta m_{\rm ATM}^2$, $\Delta m_{\rm sol}^2$,
 $\sin^2 2 \theta_{12}
 $
 , and 
 $\sin^2 2 \theta_{23}
 $,
 the value of $U_{e3}$ remains as
 an unknown parameter, which we only know the upper bound
 from CHOOZ experiments as
 $U_{e3}< 0.16$\cite{CHOOZ}.
Therefore
 if we would like to predict the solar neutrino solutions
 from the value of $U_{e3}$,
 one additional nontrivial
 condition must be needed\cite{HS}.
In Ref.\cite{ABR},
 they have denoted $\varepsilon \equiv \alpha + \beta$
 and $\varepsilon' \equiv \alpha - \beta$, where
 $\alpha \equiv \sqrt{2} U_{e3}
 (1 - {(m_1 \cos^2 \theta_{12}+ m_2 \sin^2 \theta_{12}) / m_3})$
 and $\beta \equiv {\sqrt{2}
 (m_2 - m_1)\cos \theta_{12}\sin \theta_{12}/ m_3}$,
 and
 said that $\varepsilon + \varepsilon'$ and
 $\varepsilon - \varepsilon'$ are
 expected to be of the same order if there are no
 accidental cancellations.
However,
 $\alpha$ is the free parameter which
 has nothing to do with $\beta$ at all.
Therefore, the condition $\varepsilon + \varepsilon' \simeq
 \varepsilon - \varepsilon'$ means $\alpha \simeq \beta$,
 which is not the natural condition
 but just the trivial assumption.

What are considerable nontrivial conditions from the theoretical
 point of view?
How about introducing the new
 lepton number symmetry conservation, for example,
 $L_{new}\equiv L_{e}-L_{\mu}-L_{\tau}$\cite{leptonN}?
It is useful
 only for the case of $m_1 \sim m_2 \gg m_3$
 with bi-maximal mixing (case(b2))\cite{type}\cite{HO1}.
In this case the $L_{new}$
 symmetric mass in (1,2) and (1,3) elements
 in the neutrino mass matrix are leading,
 and other $L_{new}$
 symmetry breaking elements are negligibly small.
If
 elements of (1,2) and (1,3) are the same order
 including the $L_{new}$ symmetry breaking order,
 we can obtain the
 relation between $U_{e3}$ and $\sin \theta_{12}$ as
\begin{equation}
 U_{e3} \simeq (1- \sin 2 \theta_{12})
   \left(\frac{m_3 }{ m_1}-\frac{|m_2|-|m_1|}{ 2 m_1}\right)^{-1}.
\end{equation}
However, the
 equality of (1,2) and (1,3) elements
 including the $L_{new}$ symmetry breaking order
 is the trivial assumption again.
In other cases of mass hierarchies of
 $m_1 \ll m_2 \ll m_3$ and $m_1 \sim m_2 \gg m_3$,
 the relations induced from the condition
 that the elements of (1,2) and (1,3)
 are the same order\cite{ABR} are also trivial\cite{HS},
 since there are no physical reason.

{}For another theoretical
 nontrivial condition,
 how about considering
 the situation of
 the maximal mixing between
 the second and the third generations and
 the zero mixing between the first and the third
 generations at the high energy scale?
In this case
 the quantum corrections
 can change the mixing angles
 and mass squared differences\cite{HO1}.
Let us pick up the case of
 $m_1 \sim -m_2 \sim m_3$ (case (c2) in Ref.\cite{HO1}),
 and estimate the quantum effects.
When the effect of quantum corrections is
 large enough to
 neglect neutrino
 mass differences,
 the mixing angles of the MNS matrix approach
 to certain fixed angles\cite{HO1}.
In the large limit of
 quantum collections
 the relation between the values of $U_{e3}$ and
 $U_{e2}$ is obtained as
\begin{equation}
\label{2}
 U_{e3} = U_{e2}\;
  \sqrt{1- \; U_{e2}^2 \over 1+ \; U_{e2}^2}
\end{equation}
at the low energy
 independently of the mass squared differences.
This determines the values of $U_{e3}$ according to the
 values of mixing angles of the
 solar neutrino solutions,
 which are given by
\begin{equation}
 U_{e3} \simeq 0.4 \;\;\;({\rm large \; angle}),\;\;\;
               0.016 \;\;\;({\rm small \; angle}).
\end{equation}
This means that the large angle solutions are not
 consistent with the CHOOZ experiment.
Also in the small angle solution,
 the mass squared differences
 $\Delta m_{\rm sol}^2 \simeq m_1^2 \eta$ and
 $\Delta m_{\rm ATM}^2 \simeq
 {\rm Max}\; [m_1^2 \eta^2, \; m_1^2 \eta \sin^2 \theta_{12}]$,
 can not suggest the suitable MSW solution
\cite{HO1}
,where
\begin{equation}
\eta \simeq \frac{1}{8\pi^2} \frac{m_\tau}{v^2}(1+\tan ^2 \beta
) \ln \left( \frac{m_h}{m_z}  \right).
\end{equation}



Here we have suggested two theoretical conditions
 in order to obtain the relation between $U_{e3}$ and $\sin \theta_{12}$.
However, in the first condition
 the equality of symmetry breaking elements
 must be needed, which is trivial assumption.
In the second condition, we cannot obtain suitable mass squared difference.
Thus, in order to obtain the nontrivial relation
 between the solar neutrino solutions and
 the value of $U_{e3}$, there must be another
 nontrivial
 condition.  \\

\par
\noindent
N. Haba and Tomoharu Suzuki \\
\hspace*{.5cm}Department of Physics, Nagoya University, \\
\hspace*{.5cm}Nagoya, Japan 464-8602 \\
\vspace{-1cm}

%
%

\end{document}